\documentclass[pra,twocolumn,aps,showpacs,tightenlines,superscriptaddress,amsmath,amssymb,amsfonts]{revtex4}
\usepackage{amssymb,amsmath,amsthm,amsbsy,epsfig,color,graphicx,times}%,bbm}
\usepackage[ansinew]{inputenc}
\usepackage[english]{babel}
\usepackage{accents}
\usepackage{color}
\usepackage{braket}
\usepackage{booktabs}
\usepackage{tabularx}
\usepackage{array}
\usepackage{hyperref}
\usepackage{bbold}

\begin{document}

\title{Geometric phase of quenched spin models}

\author{G. Zonzo}
\affiliation{Dipartimento di Fisica ``E.R.Caianiello'' Universit\`a degli Studi di Salerno,
Via Ponte don Melillo 1, I-84084 Fisciano (SA), Italy}
\affiliation{INFN gruppo collegato di Salerno, Via Ponte don Melillo 1, I-84084 Fisciano (SA), Italy}
\author{A. Capolupo}
\affiliation{Dipartimento di Fisica ``E.R.Caianiello'' Universit\`a degli Studi di Salerno,
Via Ponte don Melillo 1, I-84084 Fisciano (SA), Italy}
\affiliation{INFN gruppo collegato di Salerno, Via Ponte don Melillo 1, I-84084 Fisciano (SA), Italy}
\author{S. M. Giampaolo}
\affiliation{International Institute of Physics, UFRN, Anel Vi\'ario da UFRN, Lagoa Nova, 59078-970 Natal - RN, Brazil}

% \author{V. E. Korepin}
% \affiliation{C.N. Yang Institute for Theoretical Physics, Stony Brook University, Stony Brook, NY 11794-3840, US}

\begin{abstract}
% We consider the geometrical phase of a single spin state associated to the evolution induced by the sudden quench of the set of the Hamiltonian 
% parameters of a many body system in which the spin lives.  
We consider a one dimensional spin-$1/2$ many body systems which initial state is a symmetry broken ground state and in which an evolution is induced
by a sudden quench of the Hamiltonian parameters. 
We show that the long-time behavior of the spin state, can be approximated by the one of an open two level system in which the evolution 
preserves all the symmetries of the Hamiltonian.
% 
% We consider a spin belonging to a many body system in a magnetically ordered phase, which initial state is a symmetry broken ground state. 
% We assume that in this system a sudden quench of the Hamiltonian induces an evolution. 
% We show that the long time behavior of the spin state, can be approximated by the one of an open two level system in which the evolution 
% preserves all the symmetries of the Hamiltonian.
Exploiting such a result we analyze the geometric phase associated with the evolution of the single spin state and we prove analytically that its 
long-time behavior depends on the physical phase realized after the quench. 
When the system arrives in a paramagnetic phase, the geometric phase shows a periodicity that is absent in the other cases. 
Such a difference also survives in finite size systems until boundary effects come into play.
We also discuss the effects of a explicit violation of the parity symmetry of the Hamiltonian and possible applications to the 
problem of the entanglement thermalization. 
\end{abstract}

\maketitle

The geometric phase~\cite{Pancharatnam1956,Berry1984,Aharonov1987,Samuel1988,Shapere1989} appearing in the evolution of many physical systems has 
attracted an increasing interest in the recent years.
It is related to geometrical properties of the operator which induces the evolution, and hence it provides informations about its properties.
Therefore it is not a surprise that it has been subject of several theoretical~\cite{Falci2000,Carollo2003,Carollo2004} and experimental 
investigations~\cite{Zhang2005,Leek2007,Pechal2012,Murakawa2013}.
Applications of the geometric phases range from neutrino physics, where it can provide the control on the entanglement detection by means of violation 
of Bell-like inequalities~\cite{Bertlmann2004} and can be associated to the particles 
oscillations~\cite{Capolupo2011,Capolupo2009,Capolupo2015,Capolupo2016,Johns2017}, to the NMR based quantum computation~\cite{Jones2000},
and the design of temperature sensors~\cite{Capolupo:2013xza,Capolupo:2015ina}.
It was also extensively analyzed in the field of condensed matter~\cite{Korepin1991,Niu1999,Bruno2004} where several authors have used the geometrical
phase to signal and characterize the presence of quantum critical points~\cite{Carollo2005,Zhu2006,Cucchietti2010,Tomka2012} 

The present letter also analyzes the relationship between the geometric phase and the phase transitions. 
But respect to the previous works our results are a substantial step forward. 
In fact, as we show in the following for a very large class of one dimensional spin models, considering an evolution induced by sudden quench in the 
Hamiltonian parameters the long-time dynamics of the state of the single spin can be well approximated with the one of a two-level open system that 
respects the symmetry of the many body Hamiltonian.
Exploiting this equivalence we were able to prove in a very general way that the geometric phase signal the crossing of the quantum critical point. 
Moreover, the fact that in our work we focused on a single spin state, that we take into account symmetry broken ground 
states~\cite{Sachdev2000,Hamma2017} and that we prove that the results can be extended also to finite size system makes experimental verification of 
our results possible.
Furthermore the fact that we have analyzed the geometrical phase induced by sudden quench discloses the possibility to use the geometrical phase to 
gain a deeper understanding of the entanglement thermalization~\cite{Deutsch1991,Srednicki1994,Rigol2008,Yang2017}.  

We starts by considering an one-dimensional spin-1/2 system which dynamic is governed by an Hamiltonian that preserves the parity symmetry respect to 
a spin direction that we assume to be $z$ and that at $t=0$, undergoes to a sudden quench of its parameters.
We assume that, at the beginning, the system is descried by a symmetry breaking ground state of a magnetically ordered phase of the Hamiltonian.
Focusing on a single spin of the system, we show that, after a transient, its dynamic can be well approximated by the dissipative evolution 
of an open two-level system which respects the symmetries of the Hamiltonian.
By exploiting such a result we evaluate analytically the long-time behavior of the geometrical phase of the single spin state.
We show that this behavior is dependent only on the final set of parameters of the Hamiltonian.
If the Hamiltonian defined by the new set of parameters still falls in the magnetic phase, then the geometric phase does not show any periodicity that,
on the contrary, appears in the other cases.
This result, obtained in the thermodynamic limit, can be generalized in the finite size case where symmetry broken ground states can be defined
in a discrete ensemble of points~\cite{Blasone2010}.  
Among all of them, the factorization points~\cite{Giampaolo2008,Giampaolo2009,Giampaolo2010} are the only ones in which the presence of the symmetry
broken ground states does not depend on the size of the system~\cite{Rossignoli2008}.
Using the factorization points as the initial set of parameters we show that, until boundary effects come into play, the behavior of the geometric 
phase obtained in a finite size system is indistinguishable from the one obtained in the thermodynamic case.
Therefore, the geometric phase can be used as a tool to study the properties of the physical system also by using a realization with a finite number 
of elements.
Before to conclude we also briefly discuss the effects associated to an Hamiltonian term that explicitly violate the parity symmetry and the possible 
role of the analysis of the geometric phase role in the study of the thermalization in closed quantum models.

The evolution of the density matrix $\rho(t)$ describing an open two-level system is given by~\cite{Benatti1997,Benatti2000}
\begin{equation}
 \label{general_dynamic_1}
 \frac{d\rho(t)}{dt}=-\imath [H_l,\rho(t)]+\mathcal{L}[\rho(t)] \;.
\end{equation}
The first term in the r.h.s. of eq.~(\ref{general_dynamic_1}) represents the standard quantum mechanical evolution, induced by an Hamiltonian
acting on the spin that, in general, can be written as $H_l=\sum_\mu h_\mu \sigma_\mu$ where $\sigma_\mu$ are the Pauli operators and 
$\mu=0, \; x,\; y,\; z$ ($\sigma_0=\mathbb{1}$).
$\mathcal{L}[\rho(t)]$, is a linear map which satisfies the conditions of complete positivity and trace preservation of $\rho(t)$.
It is useful to expand  eq.~(\ref{general_dynamic_1}) in the basis of the Pauli operators. 
Associating to the reduced density matrix $\rho(t)$ a vector $\bar{\rho}(t)$ whose components are the expectation values of the $\sigma_\mu$, i.e. 
$\bar{\rho}_\mu(t)=\mathrm{Tr}(\rho(t) \cdot \sigma_\mu)$, the eq.~(\ref{general_dynamic_1}) becomes
\begin{equation}
 \label{general_dynamic_2}
 \frac{d\bar{\rho}(t)}{dt}=-2 U \bar{\rho}(t) \;, 
\end{equation}
where $U$ is a real matrix equal to
\begin{equation}
 \label{general_dynamic_3}
 U=\left(
 \begin{array}{cccc}
  0 & 0 & 0 &0 \\
  0 & \lambda_x & \alpha - h_z & \beta - h_y  \\
  0 & \alpha + h_z & \lambda_y & \delta - h_x  \\
  0 & \beta + h_y & \delta + h_x & \lambda_z
 \end{array}
 \right) \;.
\end{equation}
%with $\lambda_\mu \ge 0\, \forall \, \mu$.

The matrix $U$ in eq.~(\ref{general_dynamic_3}) describe a completely general evolution.
However the evolution induced by a sudden quench of the Hamiltonian parameters of a many-body system preserves the parity.
In fact if a state before the quench has a defined parity, then the parity must be preserved also during the evolution~\cite{Giampaolo2017}.  
At the level of a single spin, ground states with a defined parity are characterized by the fact that $\rho_x(0)=\rho_y(0)=0$.
Hence, the operator $U$  has to satisfy the condition that if $\rho_x(0)=\rho_y(0)=0$ then $\rho_x(t)=\rho_y(t)=0 \; \forall \, t\ge0$.
This implies that $\beta=\delta=h_y=h_x=0$.

Using these assumptions we solve eq.~(\ref{general_dynamic_2}) and we obtain an explicit expression for $\rho_\mu(t)$
\begin{eqnarray}
\label{solution_differential_equation}
 \rho_x(t)&=&e^{-t\lambda_s}\left[\rho_x(0) \cosh(\omega t)- A_x \sinh(\omega t))\right] \; ; \nonumber \\
 \rho_y(t)&=&e^{-t\lambda_s}\left[\rho_y(0) \cosh(\omega t)+ A_y \sinh(\omega t))\right] \; ; \nonumber \\
\rho_z(t)&=&e^{-2 t \lambda_z} \rho_z(0) \; .
\end{eqnarray}
Here $\omega\!\!=\!\!\sqrt{4\! (\alpha^2\!-\!h_z^2)\!+\!\lambda_d^2}$, $\!\lambda_s\!=\!\lambda_x\!+\!\lambda_y$ and 
\mbox{$\!\lambda_d\!=\!\lambda_x\!-\!\lambda_y$}, while the parameters $A_x$ and $A_y$, assuming $\zeta_\pm=\alpha \pm h_z$, are %respectively equal to 
$A_x\!=\!(\!\rho_x(0)\! \lambda_d\!+\!2 \zeta_-\rho_y(0)\!)\!/\!\omega$ and
$A_y\!=\!(\!\rho_y(0)\! \lambda_d\!-\!2 \zeta_+\rho_x(0)\!)\!/\!\omega$.%, with $\zeta_\pm=\alpha \pm h_z$.
%The complete positivity of the reduced density matrix for any $t$ is guaranteed by the assumption $\lambda_s > \mathrm{Re}(\omega)$.

To show that eqs.~(\ref{solution_differential_equation}) can describe the dynamic of the state of a spin in a many body system  
which Hamiltonian undergoes to a sudden quench, we compare them with several samples in which we have analyzed the exact dynamics.
These samples are obtained by considering a thermodynamic one-dimensional system, which Hamiltonian is
\begin{equation}
\label{eq:XYmodelhamiltonian}
 H \! \! =\! \!-\!\sum_{i=1}^{N}\! \gamma_x  \sigma_i^x \sigma_{i+1}^x \!+\!
\gamma_y \sigma_i^y \sigma_{i+1}^y \!+ \!\Delta \sigma^x_{i-1}\sigma^z_{i}\sigma^x_{i+1}\! +\!\! h \sigma_i^z \, ,
\end{equation}
that can be analytically solvable~\cite{Montes2011} by using the Jordan Wigner transformations~\cite{Jordan1928}.
It includes several well known models as the quantum Ising model~\cite{Lieb1961,Barouch1970,Barouch1971}, that can be 
obtained setting $\Delta=0$, and the cluster-Ising model~\cite{Smacchia2011,Giampaolo2014} that is recovered fixing $\gamma_x=0$ and $h=0$.
Moreover, it is easy to verify that, regardless the values of $\gamma_x$, $\gamma_y$, $h$ and $\Delta$, the Hamiltonian always commutes with the 
parity operator along $z$ direction i.e. $P_z=\bigotimes_i \sigma_i^z$.

\begin{figure}[t]
\begin{center}
\includegraphics[width=8.5cm]{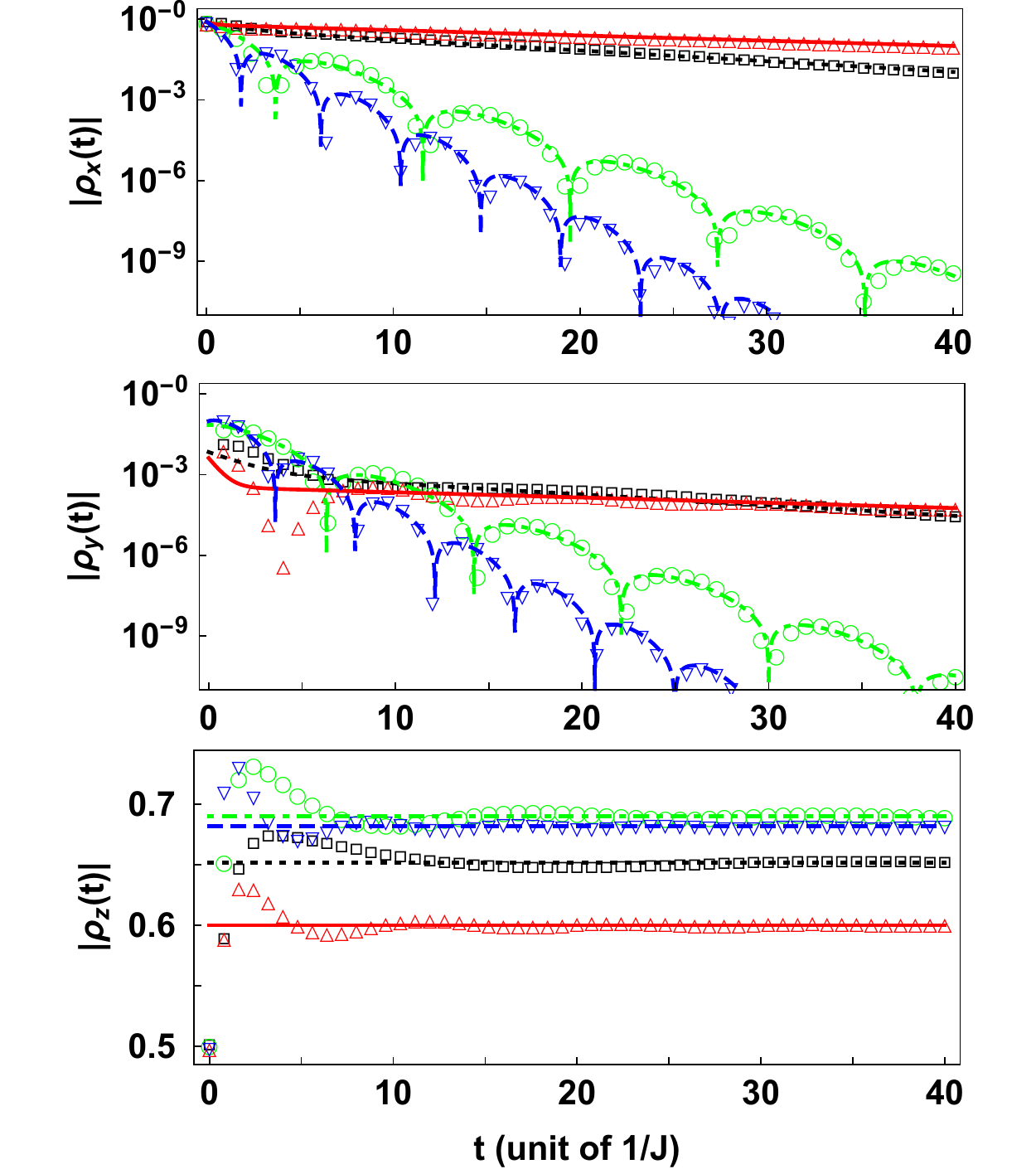}
\caption{(Color online). Behaviors of the $\rho_\mu (t)$, with $\mu=x,\, y,\, z$, as functions of the time for several different quenches.
For all the samples the dots stand for  the values obtained solving exactly the model in eq.~(\ref{eq:XYmodelhamiltonian}), by using the approach 
described in Ref.~\cite{Giampaolo2017}, while the lines are the result of a fit obtained using eqs.~(\ref{solution_differential_equation}).
In two of the quenches showed we have considered the case in which the final set of the parameter are still in the ordered phase. 
They are the black empty squares / black dotted line in which the quench starts from $\{\Delta\!=\!0,\gamma_x\!=\!0.8,\gamma_y\!=\!0.2,h\!=\!0.8\}$ 
and ends in $\{\Delta\!=\!0,\gamma_x\!=\!0.8,\gamma_y\!=\!0.2,h\!=\!0.95\}$, and the red empty up-triangles / red solid line in which the initial 
parameters are $\{\Delta\!=\!0.2,\gamma_x\!=\!0.8,\gamma_y\!=\!0.2,h\!=\!0.8\}$ and the final ones are 
$\{\Delta\!=\!0.2,\gamma_x\!=\!0.8,\gamma_y\!=\!0.2,h\!=\!0.97\}$. 
In the other two samples the final set of parameters falls in the paramagnetic region. 
For the green empty circles / green dashed lines we have that the quench starts from $\{\Delta\!=\!0,\gamma_x\!=\!0.8,\gamma_y\!=\!0.2,h\!=\!0.8\}$ 
and ends in $\{\Delta\!=\!0,\gamma_x\!=\!0.8,\gamma_y\!=\!0.2,h\!=\!1.1\}$, while the blue empty down-triangles / blue dot-dashed line in which the 
initial parameters are $\{\Delta\!=\!0.2,\gamma_x\!=\!0.8,\gamma_y\!=\!0.2,h\!=\!0.8\}$ and the final ones are 
$\{\Delta\!=\!0.2,\gamma_x\!=\!0.8,\gamma_y\!=\!0.2,h\!=\!1.3\}$. }\label{figure_1}
\end{center}
\end{figure}

In fig.~(\ref{figure_1}) we compare, for several cases, the exact results obtained for the $\rho_\mu(t)$ of the model in 
eq.~(\ref{eq:XYmodelhamiltonian}), using the approach described in Ref.~\cite{Giampaolo2017}, with a fit obtained using 
eqs.~(\ref{solution_differential_equation}).
The exact results are obtained by considering, as initial state, a symmetry-breaking ground state.
%In fact, for a ground state with a defined parity one has $\rho_x(t)=\rho_y(t)=0\, \forall\,t$.
Looking at the behavior of $\rho_z(t)$  we note that, in all the cases analyzed, as $t$ increases, it goes to a finite value, namely 
$\rho_z(\infty)$, that only accidentally is zero. 
On the contrary $\rho_{x}(t)$ and $\rho_y(t)$ always vanishes exponentially.
However, when the set of the parameters after the quench lives in the region of the disordered phase, $\rho_x(t)$ and $\rho_y(t)$ have, in addiction 
to the exponential decay, an oscillatory behavior, that is absent in the other case.
This picture is in agreement with the results obtained in the cases of the quantum-Ising model and the cluster-Ising 
models~\cite{Calabrese2011,Calabrese2012-1,Calabrese2012-2,Giampaolo2017}.
For what concern the eqs.~(\ref{solution_differential_equation}) we see that they fail to describe the short-time behavior of the $\rho_\mu(t)$ while,
increasing $t$, the agreement with the exact solutions becomes better and better. 
It is worth to note that the short-time failure is not due to the particular sets of parameters that we have used. 
On the contrary it is connected to the fact that $\rho_z(\infty)$ is, in general non zero and different from $\rho_z(0)$. 
Hence its behavior for all times cannot be explained using the third of eqs.~(\ref{solution_differential_equation}).
Therefore the best fit of the long-time behavior is obtained assuming in the eqs.~(\ref{solution_differential_equation}) $\rho_z(0)=\rho_z(\infty)$ 
and $\lambda_z=0$.
On the other hand, the long-time behaviors of $\rho_x(t)$ and $\rho_y(t)$ can be explained by assuming that $\omega^2$ changes from positive to 
negative values when the set of the parameters after the quench moves from ordered to disordered phase.

Hence the eqs.~(\ref{solution_differential_equation}) can be used to describe the long-time behavior of the single spin state and of any physical 
quantity defined on it.
With the aim of capturing the two different behaviors of the single spin state we choose to focus on the geometric phase $\Phi_g(t)$. 
In particular we use the Wang and Liu approach~\cite{Wang2013} that allows to evaluate the geometrical phase in an open quantum system with a 
nonunitary and noncyclic evolution. 
The geometrical phase $\Phi_g(t)$ is defined as the difference between the total $\Phi_{t}(t)$ and the dynamic phase $\Phi_{d}(t)$.
It can be expressed in terms of the time-dependent parameters representing the single spin state in the Block sphere, i.e:
1) the purity, i.e. the vector radius in the Block sphere  $r(t)=\sqrt{\rho_x^2(t)+\rho_y^2(t)+\rho_z^2(t)}$;
2) the polar angle defined as $\theta(t)=\cos^{-1}(\rho_z(t)/r(t))$;
3) the azimuth angle $\varphi(t)=\tan^{-1}(\rho_y(t)/\rho_x(t))$.

The total phase $\Phi_{t}(t)$ depends only on the initial and on the state at the time $t$.
In terms of these geometrical parameters it is written as
\begin{eqnarray}
\label{eq:total_phase_1}
\Phi_{t}(t)\!\! \!&\!\!&\!\!=\sum_{k=0}^{1}\frac{\sqrt{(1+(-1)^kr(t))(1+(-1)^kr(0))}}{2} \tan^{-1} \!\!\! \! \\
\! \!& \!\!& \!\!\!\!\!\!\!\!\!\!\!\left(\frac{\sin(\varphi(t)-\varphi(0))\sin\chi_{k,0}\sin\chi_{k,t}}
 {\cos\chi_{k,0}\cos\chi_{k,t}+\cos(\varphi(t)-\varphi(0))\sin\chi_{k,0}\sin\chi_{k,t}}\right) \nonumber
\end{eqnarray}
where $\chi_{k,t}=\frac{\theta(t)}{2}+k\frac{\pi}{2}$.
Assuming $\rho_z(\infty)\neq0$ and taking into account that $\lambda_z=0$ and $\rho_x(\infty)=\rho_y(\infty)=0$ it follows that 
$\theta(t) \rightarrow 0,\, \pi$ depending on the sign of $\rho_z(\infty)$.
Therefore, defining $s=\mathrm{sgn}(\rho_z(\infty))$, we have that, for $t \rightarrow \infty $ $\Phi_t(t)$ becomes
\begin{equation}
\label{eq:total_phase_2}
\Phi_{t}(t)\simeq\frac{1}{2}\sqrt{(1-s\, r(\infty))(1-s\, r(0))} (\varphi(t)-\varphi(0))\;,
\end{equation}
where $r(\infty)=\lim_{t\rightarrow \infty} r(t)=|\rho_z(\infty)|$.

On the contrary, the dynamic phase $\Phi_d(t)$ depends on the path followed by the state in its evolution. 
In terms of the parameters of the Block sphere it can be written as
\begin{equation}
\label{eq:dynamic_phase_1}
\Phi_d(t)=\frac{1}{2}\sum_{k=0}^1 \int_0^t (1+(-1)^kr(t')) \cos^2\chi_{k,t'} d\varphi.
\end{equation}
Having only the expression for the for the long-time dynamic of $\rho_\mu(t)$ the dependence of the dynamic phase on all the time smaller that $t$ 
can be a problem. 
However, defining $t^*$ the time in which we can start to approximate the true $\rho_\mu(t)$ with the the eqs.~(\ref{solution_differential_equation}) 
and the linearity of the integral, we can consider the dynamical phase $\Phi_d(t)$ as the sum of two terms:
the \mbox{short-time} term (between $t=0$ and $t=t^*$) which we indicate with  $\Phi^{(s)}_d$ and the long-time term (for times greater than $t^*$).
Notice that the short-time term $\Phi^{(s)}_d$ represents a constant that affects the precise value of the phase, but not its behavior.
In the evaluation of the long-time term, assuming $\rho_z(\infty) \neq 0$ we can consider $r(t) = |\rho_z(\infty)|$ and $\theta(t) = 0,\, \pi$. 
By exploiting these assumptions it is easy to see that the dynamical phase becomes
\begin{equation}
\label{eq:dynamic_phase_2}
\Phi_d(t) \simeq \Phi^{(s)}_d + \frac{1}{2} (1-s\,r(\infty)) (\varphi(t)-\varphi(t^*))
\end{equation}
% 
% Being $\Phi_g(t) = \Phi_t(t) -\Phi_d(t) $, we now analyze separately the total and the dynamic phase.
% We write them in terms of the 
% 
From eq.~(\ref{eq:total_phase_2}) and eq.~(\ref{eq:dynamic_phase_2}) it follows that, in the long-time regime, the geometrical phase
$\Phi_g(t)=\Phi_{tot}(t)-\Phi_d(t)$ depends linearly on the azimuthal angle $\varphi(t)$
\begin{equation}
 \label{geometrical_phase}
 \Phi_g(t)\simeq \frac{1}{2}C \, \varphi(t) + \mathrm{cost} \,,
\end{equation}
where $C=\sqrt{1-s\, r(\infty))(1-\,s r(0))}-(1-s\, r(\infty))$.

\begin{figure}[t]
\begin{center}
\includegraphics[width=8.5cm]{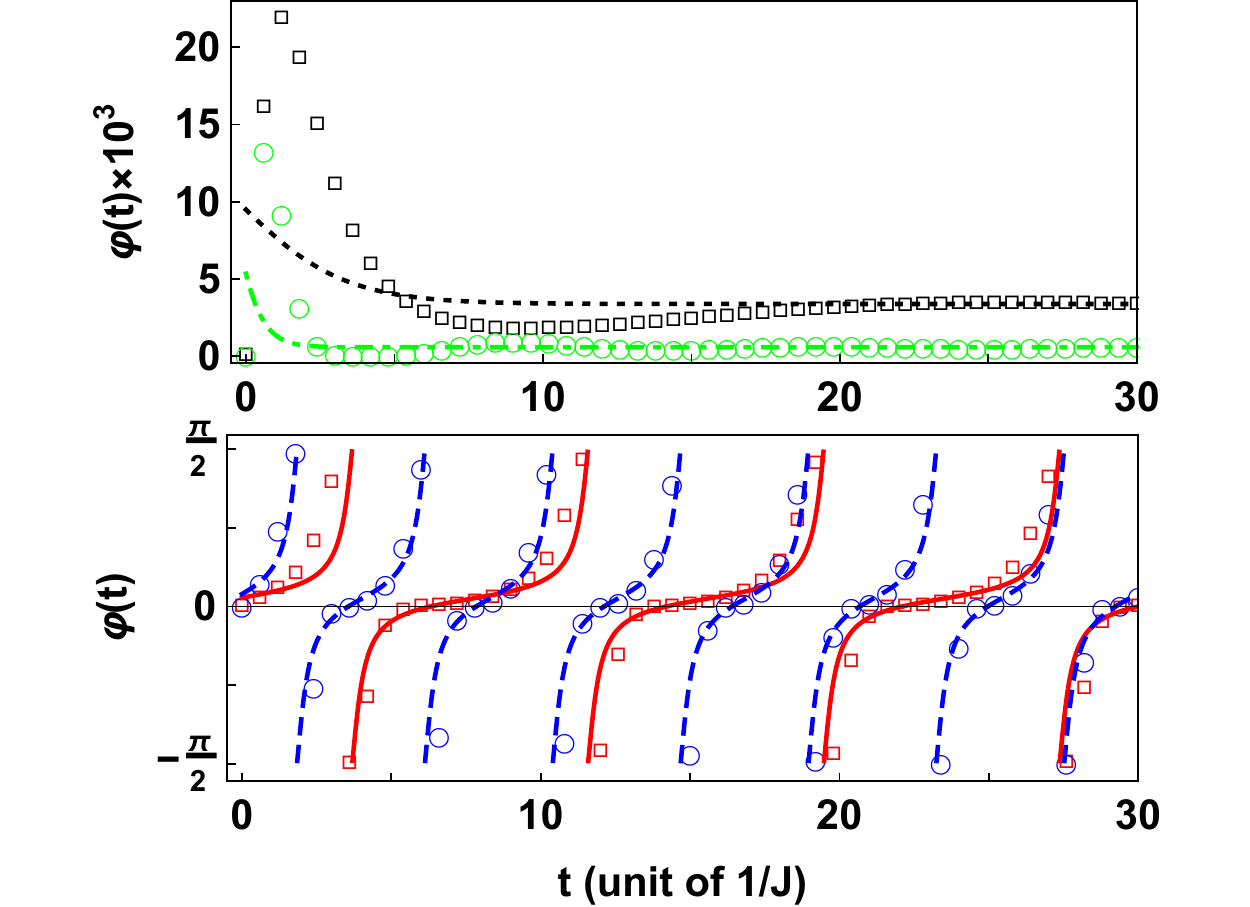}
\caption{(Color online). Behaviors of the $\varphi(t)$ as functions of the time for the cases analyzed in fig.~(\ref{figure_1}).
At the same set of parameters is associated the same code of point/line of fig.~(\ref{figure_1}. }
\label{figure_2}
\end{center}
\end{figure}

The azimuthal angle $\varphi(t)$ shows two different long-time behaviors.
If the system, after the quench, is still in the magnetically ordered phase, then, $\omega^2>0$ and $\varphi(t)$ goes to a constant value 
for $t \rightarrow \infty$. 
Consequently $\Phi_d(t)$, $\Phi_{tot}(t)$ and $\Phi_g(t)$ become constants.
On the contrary if the system has crossed the quantum critical point we have $\omega^2<0$ and $\varphi(t)$ 
becomes a periodic function of the time, with a periodicity equal to $\mathrm{Im}(\omega)$.
In this case also $\Phi_d(t)$, $\Phi_{tot}(t)$ and $\Phi_g(t)$ start to show a periodic behavior that is absent in the previous case.
In  fig.~(\ref{figure_2}) we show for the cases analyzed in fig.~(\ref{figure_1}) the behavior of $\varphi(t)$ evaluated using the 
eqs.~(\ref{solution_differential_equation}) and the exact analytical solutions of the models in eq.~(\ref{eq:XYmodelhamiltonian}).

\begin{figure}[t]
\begin{center}
\includegraphics[width=8.0cm]{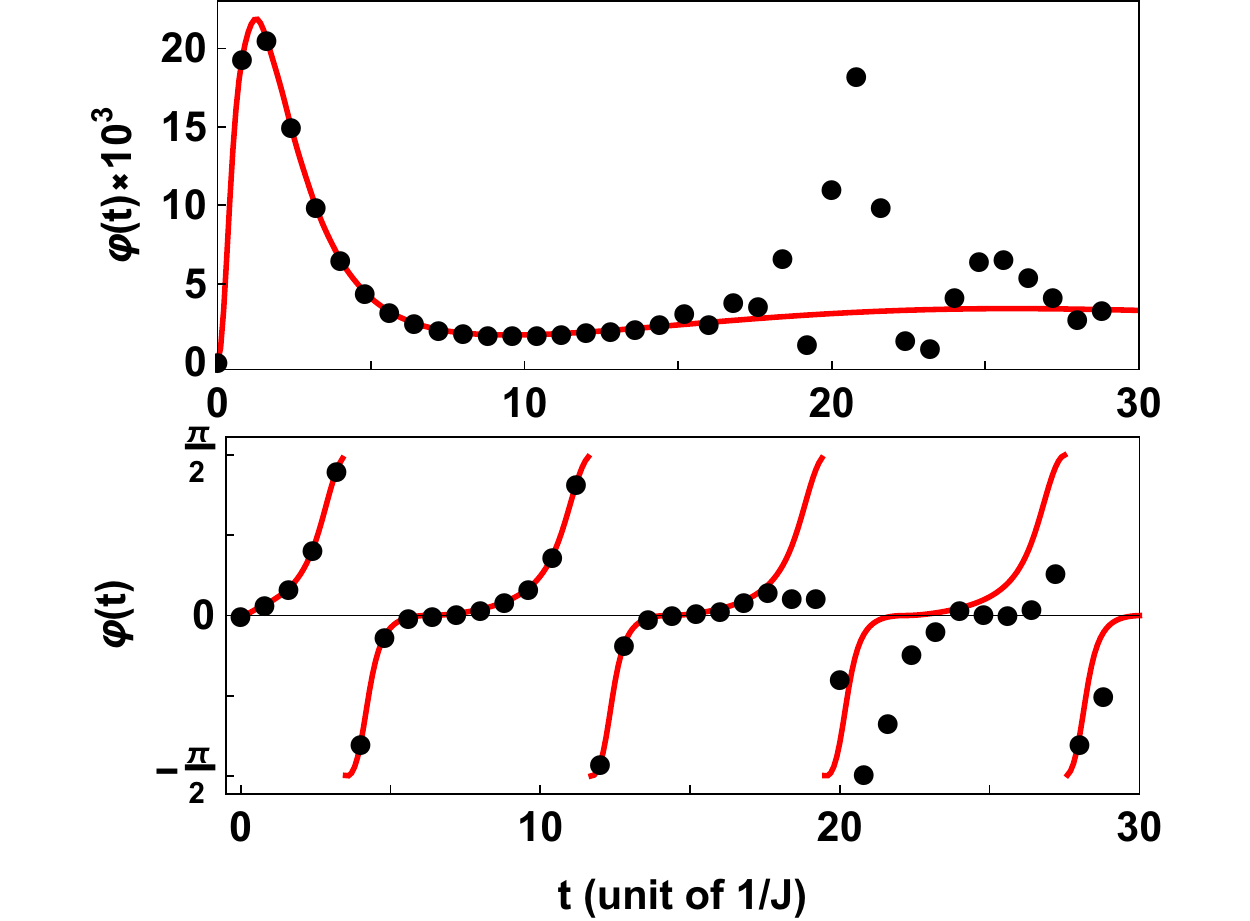}
\caption{(Color online) Finite size effects.
Behaviors of the $\varphi(t)$ associated to the evolution of the symmetry-breaking ground state for 
$\{\Delta\!=\!0,\gamma_x\!=\!0.8,\gamma_y\!=\!0.2,h\!=\!0.8\}$ fro sudden quenches to the sets of parameters 
$\{\Delta\!=\!0,\gamma_x\!=\!0.8,\gamma_y\!=\!0.2,h\!=\!0.95\}$ (upper panel) and $\{\Delta\!=\!0,\gamma_x\!=\!0.8,\gamma_y\!=\!0.2,h\!=\!1.1\}$.
The initial set of parameters is a factorization point and hence the superposition ground state exists also at finite size.
In both the two panel the black dots stand for the results obtained with a system made by 100 spins while the red line are the data obteined 
in the thermodynamic limit.}\label{figure_3}
\end{center}
\end{figure}

Let us now take into account the particular case in which \mbox{$\rho_z(\infty)=0$}.
In this situation, from the definition of the parameters of the Block sphere, we have that $\theta(t)= \frac{\pi}{2} \, \forall \, t$ 
and for $t\rightarrow\infty$, one has $r(t) \rightarrow 0$.
For the dynamical phase, following the same approach used in the general case we obtain a result that is equivalent to the general case, i.e.
\begin{equation}
\label{eq:dynamic_phase_3}
\Phi_d(t) \simeq \Phi^{(s)}_d + \frac{1}{2} (\varphi(t)-\varphi(t^*))
\end{equation}
On the contrary the total phase $\Phi_t(t)$ shows an expression completely different from  eq.~(\ref{eq:total_phase_2}).
Indeed being $\theta(t)= \frac{\pi}{2} \, \forall \, t$ the total phase, in the limit of large times, becomes
\begin{eqnarray}
\label{eq:total_phase_3}
\Phi_{t}(t)\!\! \!&\!\!&\!\!=\frac{1}{2} K \tan^{-1} \left(\frac{\sin(\varphi(t)-\varphi(0))}
 {1+\cos(\varphi(t)-\varphi(0))}\right) \nonumber
\end{eqnarray}
where $K=\sqrt{1+r(0)}+\sqrt{1-r(0)}$.
Therefore $\phi_t(t)$ is not more a linear function of $\varphi(t)$. 
But, also in this case, when the final set of the Hamiltonian parameters cross the quantum critical point, the total and the geometrical phase, 
pass from a constant to a periodic long-term behavior.
% Summarizing we may claim that the single spin geometrical phase $\phi_g(t)$, induced by a  
% sudden quench can be a way to analyze the physical phase that characterize the system at the end of the quench.  

The analysis of the single spin geometric phase presented here holds not only in the thermodynamic limit, but also at finite size. 
In fact also at finite size, in the region of parameter in which, at the thermodynamic limit, we have a magnetic phase, there exist several sets of 
the Hamiltonian parameters in which the system shows a degenered ground state~\cite{Blasone2010}.
Then, it is possible to construct a symmetry-breaking ground state also at finite size.
Among the sets of Hamiltonian parameters, a crucial role is played by the set of factorization points~\cite{Giampaolo2008,Giampaolo2009,Giampaolo2010},
in which the presence of the degeneracy does not depends on the size of the system~\cite{Rossignoli2008}.
In fig.~(\ref{figure_3}) we report the behavior of $\varphi(t)$ for the quenches analyzed in fig.~(\ref{figure_1}) which initial set of the 
Hamiltonian parameters coincide with a factorization point.
We note that, the finite size systems show the same behavior of the thermodynamic one, up to a certain time $t$  in which the boundary effects enter 
into play. 

The results obtained depend on the assumption that in the evolution the parity is preserved.
If we modify such hypothesis by introducing, for example, a small local field along the $x$ direction, i.e. $h_x \neq 0$ in 
eq.~(\ref{general_dynamic_3}), the picture changes suddenly.
In fact the periodic behavior of the geometric phase, for a quench crossing the critical point, is due to the
fact that $\rho_x(t)$ and $\rho_y(t)$ oscillate around $0$ (see fig.\ref{figure_1}).
In the case in which $h_x \neq 0$, the oscillations of $\rho_x(t)$ and $\rho_y(t)$ are no more centered around $0$ but around a different values.
Therefore, there exists a certain time $t$ after which none oscillations will take pace and the long-time periodic behavior of $\varphi(t)$ disappears.
Since the geometrical phase is strongly depending on the symmetries of the Hamiltonian then, the analysis of such a phase can disclose a new way to 
analyze the entanglement thermalization that it is strongly dependent on the integrability of the Hamiltonian.

In summary, we have seen that the study of the single spin geometric phase, induced by a quench in the many body system in which the spin lives, may 
allow to study the phase diagram of such system.
Starting from a ground state that breaks the parity symmetry of the Hamiltonian, we show that the long-time behavior of the single spin geometrical 
phase is depending on the phase which characterizes the system after the quench.
Such behavior can be determined analytically by considering the fact that, in the long-time regime, the evolution of the single spin in the many-body 
system is well approximated by the one of a two-level open system whose evolution preserves the parity of the state.
Exploiting this result we have shown that, when the system, during the quench, cross the quantum critical point, the long-time regime of the geometric 
phase has a periodic behavior that is completely absent in the other case
This kind of analysis can be extended also in the finite size system. 
Our results are strongly dependent on the fact that, during the evolution, the parity is preserved. 
In the case that this assumption is violated the geometrical phase show a completely different behavior.
This fact disclose the possibility to use the long-time behavior of the geometrical phase to analyze the entanglement thermalization. 
This analysis will be the subject of an our future work.

\end{document}